\documentclass[fleqn,usenatbib]{mnras}

\usepackage{newtxtext,newtxmath}
\usepackage[T1]{fontenc}
\usepackage{ae,aecompl}

\usepackage{graphicx}	% Including figure files
\usepackage{amsmath}	% Advanced maths commands
\usepackage{amssymb}	% Extra maths symbols

\title[XXXX]{Are all fast radio bursts repeating sources?}

\author[M.~Caleb et al.]
{M.~Caleb$^{1}$\thanks{Email: manisha.caleb@manchester.ac.uk},
B.W.~Stappers$^{1}$,
K.~Rajwade$^{1}$,
C.~Flynn$^{2,3}$
\\
$^{1}$ Jodrell Bank Centre for Astrophysics, School of Physics and Astronomy, The University of Manchester, Manchester M13 9PL, UK \\
$^{2}$ Centre for Astrophysics and Supercomputing, Swinburne University of Technology, P.O. Box 218, Hawthorn, VIC 3122, Australia\\
$^{3}$ ARC Centre of Excellence for All-sky Astrophysics (CAASTRO)}

\date{Accepted XXX. Received YYY; in original form ZZZ}

\pubyear{2018}

\begin{document}
\label{firstpage}
\pagerange{\pageref{firstpage}--\pageref{lastpage}}
\maketitle

\begin{abstract}
We present Monte-Carlo simulations of a cosmological population of repeating fast radio burst (FRB) sources whose comoving density follows the cosmic star formation rate history. We assume a power-law model for the intrinsic energy distribution for each repeating FRB source located at a randomly chosen position in the sky and simulate their dispersion measures (DMs) and propagation effects along the chosen lines-of-sight to various telescopes. In one scenario, an exponential distribution for the intrinsic wait times between pulses is chosen, and in a second scenario we model the observed pulse arrival times to follow a Weibull distribution. For both models we determine whether the FRB source would be deemed a repeater based on the telescope sensitivity and time spent on follow-up observations. 
We are unable to rule out the existence of a single FRB population based on comparisons of our simulations with the longest FRB follow-up observations performed. We however rule out the possibility of FRBs 171020 and 010724 repeating with the same rate statistics as FRB 121102 and also constrain the slope of a power-law fit to the FRB energy distribution to be $-2.0 < \gamma <-1.0$. 
% Localisation, along with the association of an FRB source with a counterpart at another wavelength will be crucial to the understanding of the existence of multiple populations.
All-sky simulations of repeating FRB sources imply that the detection of singular events correspond to the bright tail-end of the adopted energy distribution due to the combination of the increase in volume probed with distance, and the position of the burst in the telescope beam.
\end{abstract}


%%%%%%%%%%%%%%%%%%%%%%%%%%%%%%%%%%%%%%%%%%%%%%%%%%

%%%%%%%%%%%%%%%%% BODY OF PAPER %%%%%%%%%%%%%%%%%%

\section{Introduction}

Fast radio bursts (FRBs) are intense ($\sim$ Jy) radio flashes of millisecond duration characterised by dispersion measures (DMs; 176 to 2596 pc cm$^{-3}$) that are much too large to be due to the interstellar medium (ISM) of the Milky Way, and are consequently considered to lie at extra-Galactic to cosmological distances. Currently, 52 FRBs have been discovered \citep{frbcat} of which only one has been seen to repeat \citep[FRB 121102 or the ``repeater'';][]{nat_spitler, Scholz}, allowing for the source to be localised to a dwarf galaxy at $z \sim 0.2$ \citep{Chatterjee, SriHarsh}. Despite the deep multi-wavelength observations following the localisation, the nature of the source producing the FRB remains open to speculation \citep{Scholz,Michilli}. With the exception of the repeater, the spatial localisation of the FRBs on the sky is typically no better than a few to tens of arcminutes, making unambiguous association with counterparts (and a potential host galaxy) at other wavelengths 
challenging. It is presently unclear whether all FRBs repeat, despite much time spent on follow-up observations at various observatories around the world \citep[e.g.][]{Rane, nat_Shannon}. It may well be that the FRB population is comprised of two or more sub-populations similar to GRBs \citep{HoggFruchter}, but this may not be known definitively until more FRBs are seen (or not!) to repeat.
In the simplest case, the repeater could belong to a different evolutionary phase of a given source population with a different source count distribution relation compared to the non-repeating FRBs. 
% For instance, rotating radio transients to FRBs could potentially be a continuum across several orders of magnitude in luminosity. 
In order to understand the population as a whole, it is vital to pin down the source location to a few arcseconds or better upon discovery, as repeating FRBs are so rare. This is especially true if there is no 
afterglow or other associated emission at any other wavelength that might help to reveal the location with sufficient precision.

FRBs are tantalizing for 2 main reasons: (i) \textit{Their progenitors remain unknown}: many models and theories have been proposed, attempting to explain FRBs but no consensus has emerged. For most of the models the FRB redshift distribution is expected to track the cosmic star formation history. While their emission mechanism
is still indeterminate, their large luminosities at high inferred redshifts imply a mechanism that is much more energetic than any known source in our Galaxy. Their millisecond timescales favour neutron star progenitors which are also attractive for
producing the very high observed rate and DMs \citep{Kulkarni}. With no repetition seen in FRBs much brighter than the repeater, despite considerable follow-up effort \citep[e.g.][]{Rane} and the striking differences in rotation measures between the repeating FRB \citep[$\sim 10^5$ rad m$^{-2}$;][]{Michilli} and one-off events \citep[$\sim -220$ rad m$^{-2}$;][]{Calebpoln} the possibility of more than one FRB emission mechanism cannot be ruled out. However there is no strong evidence yet for multiple FRB populations. (ii) \textit{They are ostensibly detectable to cosmological distances}: Over the last decade, realisation has grown that FRBs could prove to be unique cosmological probes. A key observable quantity of FRBs are their DMs, which could enable us locate the ``missing baryons'' in the low-$z$ Universe as they trace for all the ionised baryons along the line-of-sight \citep{McQuinn}. The DMs of FRBs can also be combined with their rotation measures \citep{Masui, RaviSci,Michilli, Calebpoln}
to estimate the mean magnetic field of the intergalactic medium (IGM) thereby probing primordial magnetic fields and turbulence \citep{Zheng}.
FRBs might also be utilised as cosmic-rulers to provide an independent measure of the dark energy equation of state and its dependence on redshift \citep{Zhou}. The association of an FRB with a high-$z$ host galaxy is the crucial observation which would establish that FRBs can be used as cosmological probes. All three of these cosmological goals would be made easier to achieve if FRB sources were seen to repeat.

Most of the non-repeating FRBs to date have been discovered using the multi-beam receiver of the Parkes radio telescope and the single dishes of the Australian Square Kilometre Array Pathfinder (ASKAP) equipped with phased array feeds (PAFs), whose sensitivities are at least an order of magnitude less than that of the ALFA receiver of the Arecibo telescope \citep{Scholz} typically used to detect the repeater. The relatively lower sensitivities of these telescopes could mean that only the bright tail of the pulse energy distribution would be visible, leading to the detection of one-off events, depending on the luminosity function of FRBs and the distribution of repeat burst luminosities. This is supported by the fact that even the published brightest pulse from the repeater would just be detectable above the sensitivity threshold of Parkes \citep{Scholz}. 

In this paper we perform Monte-Carlo simulations of a cosmologically distributed population of repeating 
FRBs based on the observational properties of FRB 121102. In Section \ref{sec:model} we outline our simulation model and the assumptions that we adopt. The assumed distribution for the intervals between repeat pulses are described in Section \ref{sec:waittime} along with a discussion on the possibility of a FRB source producing multiple pulses being classified as a cataclysmic one-off event, and the timescale of detecting repetition during follow-up observations of FRBs using various telescopes. Finally, we present the implications of our results in Section \ref{sec:disc} and our conclusions in Section \ref{sec:conc}.

\begin{center}
\begin{table*} 
\centering
\begin{minipage}{160mm}
\caption{Specifications of the Arecibo ALFA, Parkes Multibeam, MeerKAT and ASKAP receivers.} 
\label{tab:specs}
\begin{tabular}{c c c c c c}
\hline\hline 
Parameter          & Unit                 &  Arecibo\footnote{http://www.naic.edu/alfa/gen\_info/info\_obs.shtml} &   Parkes MB  &  MeerKAT & ASKAP \\ [0.5ex] 
 &                  &    & \citep{Keith} & \citep{Camilo18} & \citep{Bannister}\\ 
\hline
Field-of-View      & $\mathrm{deg^2}$     &   30\arcmin       &   0.55          &   1.27  & 360 \\
Central beam Gain  & K $\mathrm{Jy^{-1}}$ &   11         &   0.7          &   2.7 & 0.1\\
Central beam $T_\mathrm{sys}$ & K                &   30          &   23           &   18  & 50\\ 
Bandwidth          & MHz                  &   300         &   340          &   770  & 336 \\
Frequency          & MHz                  &   1375        &   1382         &   1285 & 1320\\
Channel width      & kHz                  &   336.04  & 390.63        & 208.98  & 1000 \\
No. of polarisations  & --                &   2           &   2            & 2   & 2 \\ 
Telescope declination limit $\delta$ & deg  & $-5 \leq \delta \leq +38$ & $\leq +20$ & $\leq +44$ & $\leq +40$\\ [1ex]
\hline  
\end{tabular} 
\end{minipage}
\end{table*}
\end{center}

\section{The simulation model}
\label{sec:model}

% To date, only FRB 121102 has been seen to repeat and the extent to which it is unique remains under debate. 
In our simulations we examine the possibility of all FRB sources being repeaters and compare the results with observations. 
We only simulate repeating FRB sources and it is purely because of observational constraints and sensitivity limitations 
that some may not be observed to repeat. We examine two cases:

\begin{enumerate}
% \item In the first case, we generate similarly sized samples of repeating FRB sources to be visible at different telescopes and perform a statistical analyses of the wait-times required to detect a repeat pulse in 
% a follow-up observation. We also attempt to constrain the slope of the FRB energy distribution. 
% \item In the second case, we populate the entire sky with repeating FRB sources and determine the ratio of repeaters to non-repeaters from the pulses detectable by an observational survey at a telescope. 
\item In the first case we generate similarly sized samples of repeating FRB sources to be visible at different telescopes. The assumed intrinsic energy distribution (see below) spans the range $10^{28} - 10^{36}$ J. The lower energy cut-off was calculated by placing an FRB 121102 pulse of 
signal-to-noise (S/N)  $= 10$ at the lowest simulated redshift and determining
the energy it would need in order to retain its S/N of 10. The event rate follows a Poisson distribution, which corresponds to an exponential wait time distribution, and implies that all events are independent. We perform statistical analyses to determine the average wait-times required to detect a repeat pulse in a continuous 50 hour follow-up observation and also attempt to constrain the slope of the FRB energy distribution.

\item In the second case we once again generate similarly sized samples of repeating FRB sources to be visible at different telescopes, but with the individual pulse arrival times following a generalisation of Poissonian statistics known as the Weibull distribution as described in \cite{Oppermann}. The shape parameter of this distribution provides the temporal clustering of the pulses observed in the repeating FRB 121102. We assume a constant shape parameter of 0.34 throughout the analyses (see Section \ref{sec:waittime} for details).
The adopted energy distribution in this case spans the range $10^{30} - 10^{38}$ J. Since the pulse repeat rate estimated by \cite{Oppermann} is based on detectable pulses and specific to FRB 121102 (i.e. $z \sim 0.2$), our lower energy cut-off is based on a S/N $= 10$ pulse at the redshift of the repeater. We perform statistical analyses to determine the average wait-times required to detect a repeat pulse during a 50 hour observation split into 2 hour sessions which are randomly spaced in time.

\end{enumerate}

The simulations from \cite{Caleb_sim} of single burst FRBs have been adapted to populate the Universe with repeating FRB sources. The comoving number density distribution of these FRB sources is assumed to be proportional to the cosmic star formation history (SFH)  under the assumption that the FRB population is tied to young stars or their immediate environments. Other scenarios are certainly possible. For example, if the behaviour is tied to an external phenomenon such as plasma lensing \citep{Cordes_lensing, Main} the dependence on the star formation rate could break. While the Parkes sample of FRBs is suggestive of the population tracing the star formation rate, the ASKAP sample of FRBs has been found to be inconsistent with a population that adheres to the star formation rate with redshift \citep{Locatelli}. Further investigation is required to study the population evolution of FRB progenitors in time (see \cite{Locatelli} for a detailed analyses). 

We assume the SFH from the review paper of \cite{Hopkins} as typical of cosmic SFH measurements, which show a rise in the star 
formation rate (SFR) of about an order of magnitude between the present $(z = 0)$ and redshifts of $z \sim 2$ (see their Figure 1). We compute the product of SFH and comoving 
volume of each shell of width $dz$ as a function of $z$, and generate Monte Carlo events under this function. For simplicity, each FRB source is assumed to be radiating isotropically
with a flat radio spectrum in the absence of observational evidence of what the broadband spectrum looks like. The intrinsic energy distribution of each source is assumed to follow a power-law defined as,

\begin{equation}
N(>E) \propto E^{-\gamma}.
\label{eq:plaw}
\end{equation}

We note that the assumption of a single power-law spectrum may best reflect the intrinsic FRB emission and not the propagation effects. For instance, \cite{Cordes_lensing} suggest that plasma lenses at Gpc distances can strongly amplify pulses thereby rendering them detectable. \cite{RaviLoeb} discuss possibility of suppression of FRB emission at lower frequencies that can manifest itself as a deviation from a single power-law in the FRB spectrum. As it is still unknown what the origin of the radio emission is we do not model these effects here.
All simulated FRB pulses are normalised to the brightest pulse detected from FRB 121102 by Arecibo \citep{Marcote}. This pulse was detected with S/N $\sim 800$ and a pulse width of $W = 0.9$ ms. We estimate the isotropic energy at source to be $E = 10^{32}$ J, for redshift  $z = 0.19273$ and corresponding luminosity distance $D_\mathrm{L}(z) = 947.7$ Gpc in a standard $\Lambda$CDM cosmology. We assume the matter density $\Omega_{m} = 0.27$, vacuum density $\Omega_{\Lambda} = 0.73$ and Hubble constant $H_0 = 71$ km s$^{-1}$ Mpc$^{-1}$ \citep{Wright}.

\begin{figure*}
\centering
\includegraphics[width=7 in]{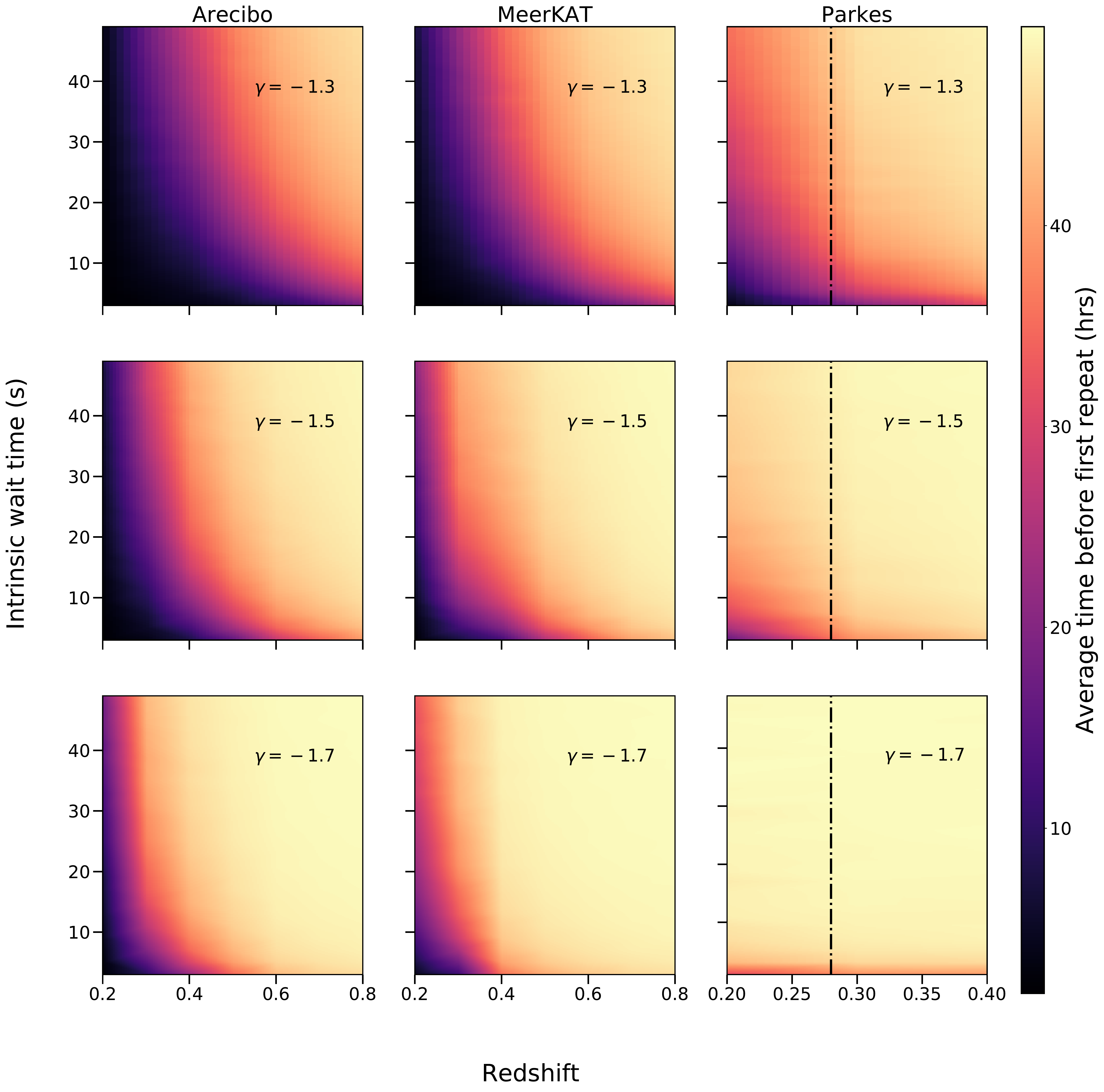}
\caption{Average, of our 1000 simulations, time in hours before detection of a repeat pulse as a function of redshift and intrinsic intervals between pulses following a Poisson distribution, for Arecibo, MeerKAT and Parkes for a total 50 hour follow-up observation. For reference, the repeater is at $z = 0.19273$. The distribution of simulated FRB pulses follows a power-law energy distribution with $\gamma = -1.3, -1.5$ and $-1.7$. The dot-dashed line represents the DM-derived maximum redshift of FRB 010724.}
\label{fig:obstime}
\end{figure*}

The total DM for any given FRB is assumed to arise from a component due to the IGM, a component due to the ISM in a host galaxy and a component due to the ISM of the Milky
Way:

\begin{equation}
\mathrm{DM_{tot}} = \mathrm{DM_{ISM}} + \mathrm{DM_{IGM}} + \mathrm{DM_{host}}.
\end{equation}

\noindent Full information on how these DM components are modelled can be found in \citet{Caleb_sim}. 
In the simulation, events are generated out to a redshift $z = 5.0$. Events are distributed randomly in the parts of the sky visible to the telescope being modelled 
(e.g.) Parkes, Arecibo, MeerKAT or the Australian Square Kilometre Array Pathfinder (ASKAP) (see Table \ref{tab:specs}). No events are generated outside the telescope horizon limits and we assume a constant time per unit sky area. While this paper has been under review, recent results from the Canadian Hydrogen Intensity Mapping Experiment (CHIME) telescope have come out \citep{NgFRB} that may allow a more detailed
look at the spectral behaviour of FRBs and also to include those detections, in particular the new repeater \citep{SriHarshFRB} in a future analysis. This inclusion however requires a good understanding of the sensitivity of the detections 
and their fluence, which is currently not well constrained \citep{SriHarshFRB}.
% a constant observation time across all declinations. 

The fluence $\mathcal{F}$ at the telescope is based on the intrinsic pulse energy $E$, the luminosity distance in the $\Lambda$CDM cosmology and a factor of $(1 + z)$ representing the time-dilation correction. We implicitly assume a power-law relationship for the energy released per unit frequency interval in the source frame. Due to the high uncertainty in spectral index of FRBs as seen in the repeater \citep{nat_spitler} we assume it to be flat (i.e. $\alpha = 0$) and there is consequently no K-correction. The fluence for such a flat spectrum is given by,

\begin{equation}
% \mathcal{F}_\mathrm{obs} = \frac{E \,(1+z)^{1+\alpha}}{4\pi \, D_\mathrm{L}^{2}(z) \, \Delta\nu} \times 10^{29} \, \, \mathrm{Jy \,ms},
\mathcal{F}_\mathrm{obs} = \frac{E \,(1+z)}{4\pi \, D_\mathrm{L}^{2}(z) \, \Delta\nu} \times 10^{29} \, \, \mathrm{Jy \,ms},
\end{equation}

\noindent where $z$ is the redshift; $D_\mathrm{L}$ is the luminosity distance in metres; $E$ is the isotropic emitted energy in J; $\Delta\nu$ is the bandwidth of the receiver system in Hz and $10^{29}$ is the 
conversion factor from Joules to  Jy ms.  
The S/N of each FRB pulse is estimated using the radiometer equation,

% \begin{equation}
% \mathrm{S/N} = \beta \, \frac {S \, G \, {\sqrt{\Delta\nu \, W \, n_\mathrm{p}}}} {T_\mathrm{rec} + T_\mathrm{sky}}
% \end{equation}

\begin{equation}
\mathrm{S/N} = \frac {\mathcal{F}_\mathrm{obs}  \, G \, {\sqrt{\Delta\nu \, n_\mathrm{p}}}} {\sqrt{W} \, (T_\mathrm{rec} + T_\mathrm{sky})}
\end{equation}

\noindent where $\mathcal{F}_\mathrm{obs}$ is fluence in Jy s, $G$ is the system gain in K Jy$^{-1}$, 
$\Delta\nu$ is the bandwidth in Hz, $W$ is the observed pulse width in seconds, ${n_\mathrm{p}}$ is the number of polarisations and ${T_\mathrm{rec}}$ and ${T_\mathrm{sky}}$ are the receiver and sky temperatures in K respectively. The sky temperature at the FRB's Galactic position $(l, b)$ is estimated from the \cite{Haslam} sky temperature map at 408 MHz. We scale the survey frequency to the telescope frequency by adopting a spectral index of $-2.6$ for the Galactic emission \citep{Reich}.

The observed width of an FRB pulse is the sum of Galactic, non-Galactic and instrumental components and can be represented as,

\begin{equation}
\label{eq:width}
W^{2} = \uptau_\mathrm{int}^{2} + \uptau_\mathrm{sc}^{2} + \uptau_\mathrm{DM}^{2}, 
\end{equation}

\noindent where $\uptau_\mathrm{int}$ is the unknown intrinsic width, $\uptau_\mathrm{sc}$ is the scatter broadening due to propagation through the interstellar medium (ISM) and intergalactic 
medium (IGM) and $\uptau_\mathrm{DM}$ is the DM smearing at the telescope. Other terms that contribute to the effective width such as the second-order correction to DM smearing, adopted 
sampling time and filter response of an 
individual frequency channel are negligible in the context of our simulations. An interesting property of the pulses from FRB 121102 is their lack of obvious pulse broadening due to 
possible multi-path scattering upon interaction with turbulent plasma along the path of propagation, quite commonly seen in pulsars and some FRBs \citep{nat_spitler, Scholz}. 
%The width of an FRB pulse is the sum of contributions from Galactic and non-Galactic components. 
The Galactic contribution to the FRB pulse width can be neglected since pulsars at similar Galactic latitudes exhibit orders of magnitude smaller scattering timescales than those seen in FRBs \citep{Krishnakumar, Bhat}.
The non-Galactic contributions to the FRB widths could arise from the host galaxy and the IGM. \cite{Macquart}  in their empirical scaling relation between
DM and scattering, show that the IGM's contribution to the pulse broadening is orders of magnitude smaller than the Milky Way's ISM. The non-monotonic dependence of pulse width on DM 
for FRBs suggests that the IGM through which all the
FRBs traverse is not responsible for the pulse broadening, which makes the host galaxy and the progenitor circum-burst medium strong candidates for the origin of the scattering. 
We thus ignore all the effects due to multi-path propagation in the simulations. 
The intrinsic pulse widths of the repeater pulses \citep[$\sim 2 - 9$ ms;][]{Hardy, Scholz} are on average observed to be much longer than most of the one-off events detected at Parkes. We note that this could be due to selection effects given that Parkes could be severely S/N limited in the population it detects compared to Arecibo. There are exceptions though, which are temporally unresolved \citep{Champion,Bhandari} and which exhibit or double- or multi-peaked pulse profiles \citep{Champion, FarahFRB}.
The wide range of observed intrinsic pulse widths makes it tough to model at present and would require further work once a larger population is made available.
The observed width is therefore estimated to be the quadrature sum of an adopted intrinsic width of 1 ms and estimated DM smearing at the telescope.

%The width of an FRB pulse affects its detection signal-to-noise ratio (S/N).
% The S/N of each simulated pulse is additionally degraded by the telescope receiver temperature and the sky temperature at the FRB's Galactic position $(l, b)$ from the \cite{Haslam} sky 
% temperature map at 408 MHz. We scale the survey frequency to the telescope frequency by adopting a spectral index of $-2.6$ for the Galactic emission \citep{Reich} and correct the S/N by a factor of $T_\mathrm{rec}/(T_\mathrm{rec} + T_\mathrm{sky})$ where $T_\mathrm{rec}$ is the telescope receiver temperature in K and $T_\mathrm{sky}$ is the sky temperature at the FRB sky position. The S/N of each pulse is then further degraded by a factor proportional to $\sqrt{W}$ where $W$ is the pulse width from Equation \ref{eq:width}. 
This $\sqrt{W}$ factor limits the redshift out to which events can be detected above the detection threshold of 10, as dispersive effects typically dominate at high redshifts. The maximum redshift we simulate is more than sufficient to sample the DM space of the published FRBs. We do not degrade the pulses by simulating a random position in the telescope beam pattern as we assume the spatial positions of the repeaters on the sky to be known, and that each repeat pulse is detected at boresight.

%\begin{figure*}
%\centering
%\includegraphics[width=7.0 in]{Fig2.pdf}
%\caption{Total number of repeat pulses as a function of redshift and the total number of repeating FRBs for Arecibo, MeerKAT and Parkes. \textbf{\color{red}Remake figure}}
%\label{fig:repeats}
%\end{figure*}

\begin{figure*}
\centering
\includegraphics[width=5.0 in]{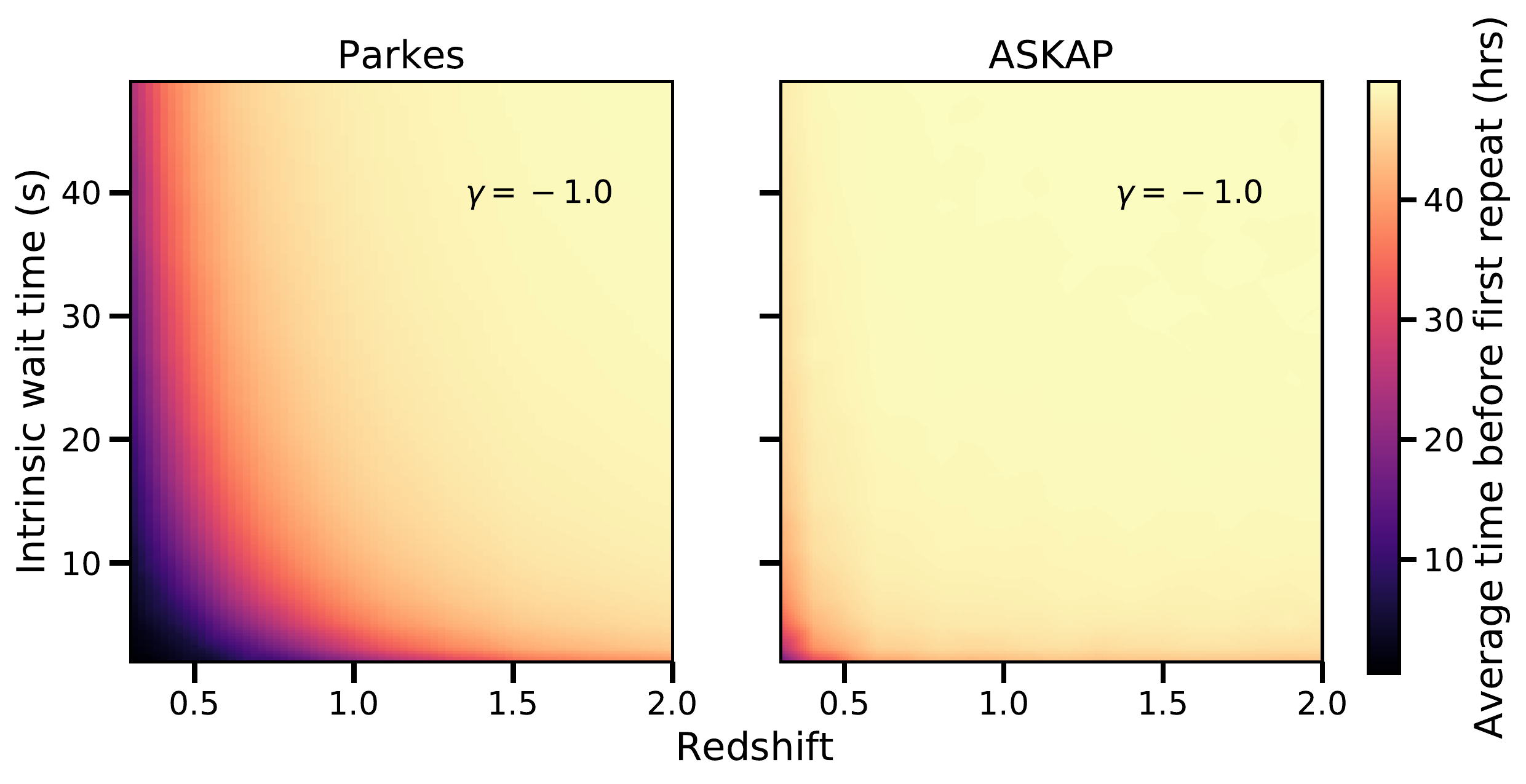}
\caption{Average, of our 1000 simulations, time before detection of a repeat pulse as a function of redshift and as a function of intrinsic intervals between pulses for the Parkes and ASKAP (incoherent) telescopes for a total 50 hour follow-up observation after an FRB is discovered. The FRB 
simulated pulses in this plot follow a power-law energy distribution with $\gamma = -1.0$.}
\label{fig:waittime1}
\end{figure*}

\section{Analysis and results}
\label{sec:waittime}

We generate 10,000 FRBs each producing $10^{6}$ pulses. The pulse energies are randomly sampled from the adopted power-law function in Equation \ref{eq:plaw} assuming slopes of $\gamma = -1.3, -1.5$ and $-1.7$. 
We classify any FRB source which results in more than one pulse detected with S/N $\geq 10$ during our simulated observation time as a repeater, and
those with only one pulse with S/N  $\geq 10$ during the same time, as one-off events.
The repeating FRB is observed to exhibit sudden periods of intense activity followed by long periods of quiescence \citep{nat_spitler,Chatterjee}. Based on these observations, we model the time intervals between pulses two different ways: 

\begin{enumerate}

\item \cite{GerryZhang} report the highest number of pulses (93 pulses) detected in a single observation (5 hours) and show that observed intervals are more
consistent with Poisson statistics than previously reported (see below), during the `on' state of the repeater. Therefore, under the assumption that an observer stays on source after the discovery of an FRB (i.e. the source is `active'), we assume an exponential model for the intrinsic intervals between the pulses. Each of the pulses is assigned a time stamp from a distribution following

\begin{equation}
% f\Big(x;\frac{1}{\beta}\Big) = \frac{1}{\beta} e^{-\big(\frac{x}{\beta}\big)},
% f(x; \lambda) = \lambda e^{-x \lambda},
% N(>\Delta t) \propto e^{-\uplambda \Delta t},
\mathcal{P}(\delta | r) = re^{-\delta r},
\end{equation}

\noindent where $\delta$ and $r$ are the expected wait time between bursts and rate respectively. The exponential is recovered from the Weibull discussed below, under the assumption of $k = 1$.
% \noindent where $\lambda$ is the rate parameter and, the scale parameter $\beta =  1/\uplambda$ and ranges from $1 \leq \beta \leq 50$ seconds. 
The model might also be valid for ``other" repeaters if the variability in FRB 121102 is
due to something extrinsic that does not affect all FRB sources. 
We note that the detection statistics in this case are affected only by total time spent on source.

\item \citet{Niels} adopt a Weibull distribution to describe the observed clustering of pulses in FRB 121102. They calculate a mean repetition rate of $ r = 5.7^{+3.0}_{-2.0}$ day$^{-1}$
above 20 mJy for a clustering parameter $k = 0.34^{+0.06}_{-0.05}$ \citep{Niels, LiamEmily}. The estimates were made based on 17 pulses from \cite{nat_spitler, Scholz, Chatterjee} detected at 1.4 and 2 GHz. We assume the same repetition rate at all the modelled telescopes. 
The probability density function of arrival times following a Weibull distribution can be described as,

\begin{equation}
\label{eq:weibull}
\mathcal{W}(\delta|k, r) = \frac{k}{\delta} \Big[\delta \, r \, \Gamma \Big( 1 + \frac{1}{k}\Big) \Big]^{k} \, e^{-\big[\delta \, r \, \Gamma \big( 1 + \frac{1}{k}\big) \big]^{k}},
\end{equation}

\noindent where $\delta$ is the distribution of intervals between subsequent bursts, and $k$ and $r$ are the shape and rate parameters as previously defined.

\end{enumerate}

The derived parameters of the Weibull distribution in \cite{Oppermann} are based on 17 pulses across 80 observations \citep{Niels, LiamEmily}. The distribution of intervals between the 93 pulses reported in \cite{GerryZhang} are observed to be more Poissonian than reported in \cite{Niels}. They however are unable to reconcile the distribution of their 15 strongest pulses with a Poissonian distribution implying that observational bias likely played a role in previously reported behaviour \citep{GerryZhang}.

% We do not model the Weibull distribution for three reasons: i) we only simulate telescopes operating at 1.4 GHz due to the variability in the spectral index of FRBs, ii) the pulses observed
% at 1.4 GHz at the Arecibo telescope were detected using different backends, iii) the simulated telescopes have varying flux thresholds, some of which are higher than 20 mJy implying a much lower repetition rate at those 
% telescopes. 
% % The statistics and repeat rates of FRBs could vary from source to source even if they are from a single progenitor class. 
% Since FRB 121102 is the only known repeating FRB, 
% it may not provide global constraints on the repeat timescales between bursts.

% \begin{figure*}
% \centering
% \includegraphics[width=7.1 in]{Fig3.pdf}
% \caption{Simulated distributions of repeating FRBs detected as one-off events at the Arecibo, MeerKAT and Parkes telescopes as a function of redshift for various values of $\gamma$.}
% \label{fig:waittime}
% \end{figure*}

\subsection{Model 1: Poisson distribution }
\subsubsection{Constraints on repeat timescales}
We first consider the case of the poissonian distribution. For the FRBs classified as repeaters, Figure \ref{fig:obstime} shows the average wait-time before the first repeat pulse would be detected above a S/N threshold of 10, as functions of 
redshift and intrinsic pulse interval for a 50 hour follow-up observation at Arecibo, MeerKAT, and Parkes given the specifications in Table \ref{tab:specs}. 
% We do not model CHIME due to its lower
% operating frequency (400 - 800 MHz) and the fact that the spectral indices of FRB pulses are highly uncertain \citep{nat_spitler}. 
Only those redshift bins with a sufficient number of repeaters per bin are considered as statistically significant and are shown in the figure.
\cite{Law121102} estimate the slope of FRB 121102's energy distribution to be $\gamma = -1.7$. If all FRBs had a similar energy distribution, for a 50 hour follow-up observation Parkes would not detect a repeater unless the intrinsic pulse interval was $< 1$ second and the source was at low redshift (see Figure \ref{fig:obstime}). For $\gamma = -1.5$ Parkes would only detect repeats if these 
FRBs had an intrinsic interval between pulses of $< 3$ seconds and were confined to $z \leq 0.35$. 
% For example, if an FRB discovered at Parkes at $z = 0.3$ were followed-up for 50 hours and no repeat was detected, we can rule out intrinsic pulse intervals greater than 3 seconds from Figure \ref{fig:obstime}. 
However as we can see from Figure \ref{fig:obstime} this does not rule out the possibility of the source being a repeater as the same source 
could be detectable as a repeater at MeerKAT or Arecibo over all the intrinsic wait timescales plotted. The comparatively higher sensitivities of Arecibo and MeerKAT not only favour detections of a large number of repeaters out to cosmological distances for flatter slopes of the intrinsic energy distribution, but also a relatively smaller number for steeper slopes (see Figure \ref{fig:obstime}) unlike Parkes.  A similar argument has been made by \cite{Trott} and \cite{Hassall}.
% {\color{red} Add MeerKAT and Arecibo stuff.}
%The average number of repeats expected per redshift bin are shown in Figure \ref{fig:repeats}.

\begin{figure*}
\centering
\includegraphics[width=7 in]{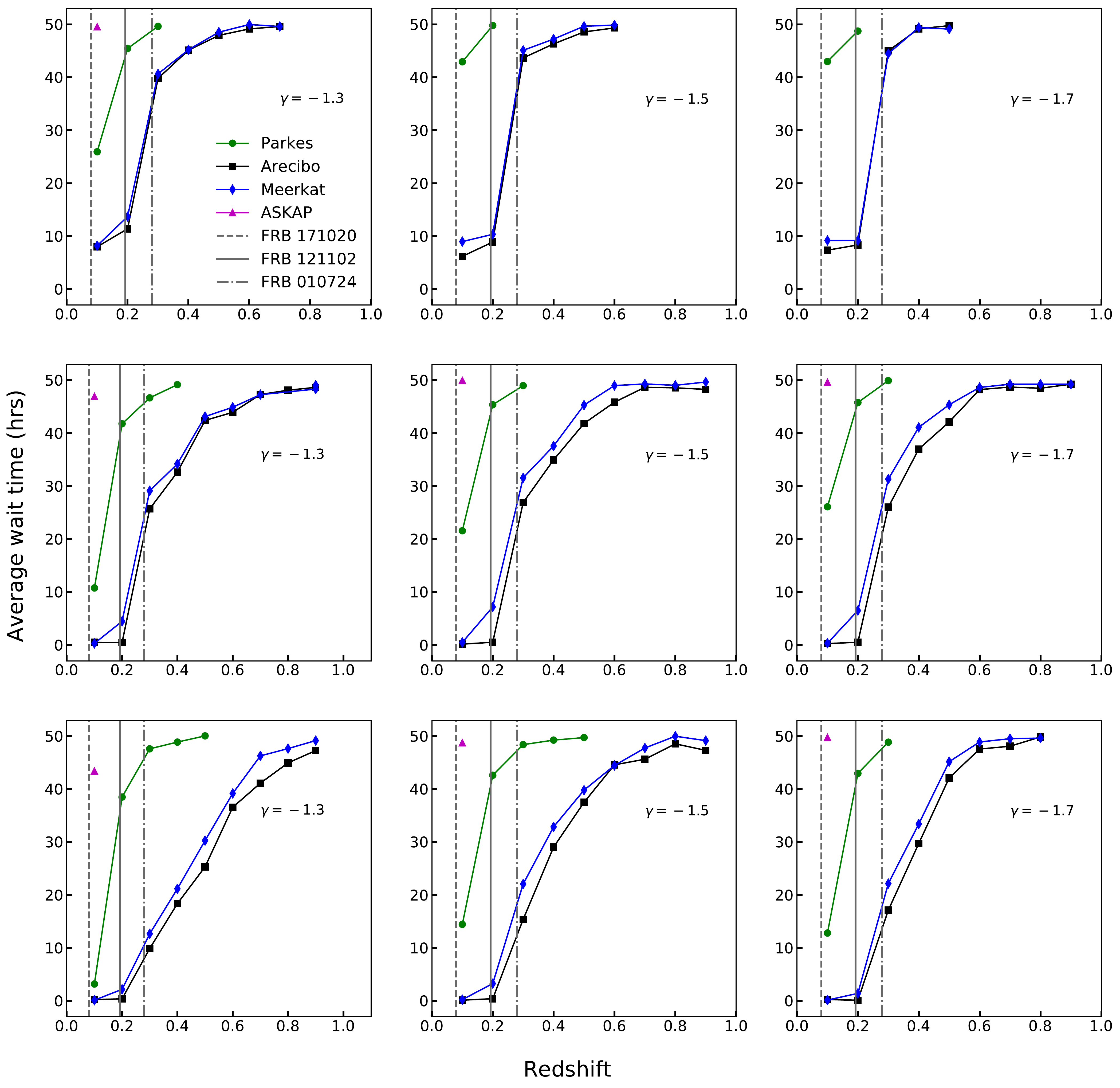}
\caption{Average time in hours before detection of a repeat pulse as a function of redshift and intrinsic intervals between pulses following a Weibull distribution, for Arecibo, MeerKAT and Parkes for a total 50 hour follow-up observation split into randomly spaced 2 hour sessions. The upper panels represent a rate of 1 pulse day$^{-1}$, the middle panels represent a rate of 5.7 pulses day$^{-1}$ and the lower panels represent a rate of 10 pulses day$^{-1}$. The slopes of the energy distributions are given by $\gamma$. The solid line represents the measured redshift of FRB 121102 while the dashed and dot-dashed lines represent the upper limits on the redshifts of FRBs 171020 and 010724 derived from their DMs.}
\label{fig:weibull}
\end{figure*}

% \subsubsection{Constraints on the slope of the energy distribution}
We attempt to constrain the slope of the energy distribution by re-running the simulations for Parkes, assuming $\gamma = -1.0$. The results are shown in Figure \ref{fig:waittime1}. 
We see that for all FRBs out to 
$z = 0.6$, given a total 50 hour observation we should detect a repeat pulse for the range of intrinsic timescales plotted. This is inconsistent with the follow-up observations of FRBs performed at Parkes.
For example, follow-up observations totalling $\sim 215$ hours were performed on FRB 150807 \citep{Rane}. From the observed DM of $266.5 \pm 0.1$ pc cm$^{-3}$,
FRB 150807 can be inferred to be at a maximum redshift of $z = 0.19$, assuming no contribution from a host galaxy and progenitor environment, with an isotropic energy of $\sim 10^{32}$ J at this 
distance \citep{RaviSci}. If FRB 150807 were similar to a simulated repeating FRB at an identical redshift, from Figure \ref{fig:waittime1} it is highly likely for a repeat pulse to have been detected.  
Similarly, FRB 010724 (a.k.a the `Lorimer burst') was
detected with a DM of $375$ pc cm$^{-3}$ and an estimated isotropic energy of $\sim 10^{33}$ J at a maximum redshift of $z = 0.28$. 
Figure \ref{fig:waittime1} suggests that even for the largest intrinsic wait time of 50 seconds, on average it would only take $\sim 25$ hours to detect a repeat pulse at this redshift. Given the $> 200$ hours \citep{Rane} spent looking for repeatability, the probability of not detecting a repeat pulse is $3.3 \times 10^{-4}$. 
This suggests that not all FRB sources might be repeatable, though this is still highly dependent on the wait timescales modelled. 
% FRB 110214 is one of the highest fluence FRBs detected to date and has an observed DM of $168.9\pm 0.5$ which places its origin at no further than $z = 0.14$ {\color{red}(Petroff et al., 2018)}. No repetition was reported in $\sim 100$ hours of follow-up observations. From our simulations the we infer the probability of non-detection to be 0.01.

The same experiment was performed for $\gamma = -2.0$ at Parkes, and resulted in an insufficient number of FRB sources (both repeating and non-repeating) to implement robust statistical analyses, implying that this slope is unlikely. The non-detection of repeat pulses from either of these FRB sources discussed, constrains the slope of the intrinsic 
energy distribution to be $-2.0 < \gamma < -1.0$ based on Figures \ref{fig:obstime} and \ref{fig:waittime1}.
A similar simulation for an ASKAP array assuming 8 dishes in an incoherent sum mode was performed. Figure \ref{fig:waittime1} indicates that given the sensitivity, it is highly unlikely that ASKAP will detect repeat pulses from FRB sources given the assumed energy range unless the slope of the energy distribution is flat and the events are at low redshift with intrinsic wait timescales shorter than one second.

\subsubsection{All-sky survey simulations}

We simulate an all-sky survey of 10,000 repeating FRBs, each producing pulses from an energy distribution (spanning $10^{28} - 10^{36}$ J) of slope $\gamma = -1.7$ and determine the number visible in the Parkes sky. An important assumption we make is that the survey is complete across all the sky visible to the telescope. In addition to the propagation effects detailed in Section \ref{sec:model} we also degrade the S/N of the FRB pulse by its randomly chosen position in the Parkes beam pattern where each receiver beam is modelled as an Airy disc with a 14.4 arcmin full-width
half-maximum. Only 18 of the simulated FRB sources are detected with S/N $\geq 10$ and declination $\delta \leq +20$, of which 15 are classified as one-off events and 3 are classified as repeaters. 
% In the Parkes simulations, 22 FRB sources are detected as one-off events and only 7 FRB sources are detected as repeaters. 
However these FRB sources
would only be detectable as repeaters if their intrinsic pulse timescales do not exceed 3 seconds as seen in Figure \ref{fig:obstime}.
We are unable to rule out a single population as it would depend highly on the unknown intrinsic repeat timescale.

\begin{figure*}
\centering
\includegraphics[width=5 in]{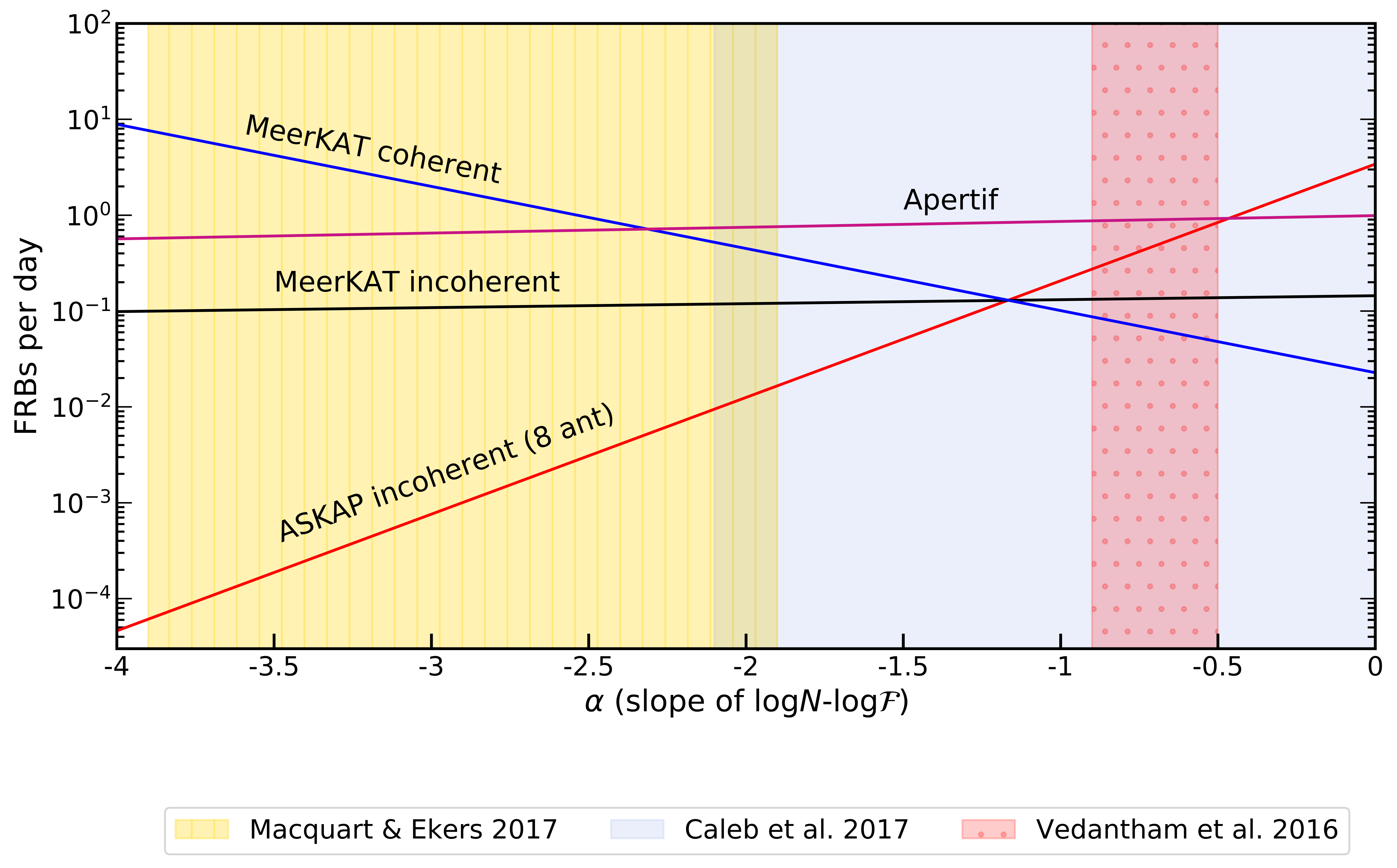}
\caption{FRB rates as a function of the slope of the integral source count distribution for various present and future FRB detection programs and telescopes. The shaded regions represent the different constraints on the slope from the literature.}
\label{fig:rates}
\end{figure*}

\subsection{Model 2: Weibull distribution}

\subsubsection{Constraints on repeat timescales}

We pick one FRB source for each simulated redshift bin of $dz = 0.1$ from our simulations and assign random arrival times to each of its $10^6$ pulses from Equation \ref{eq:weibull} using repeat rates of $r = 1, 5.7$ and 10 pulses day$^{-1}$ and a constant shape parameter $k = 0.34$ for all the modelled telescopes. The choices of rates are motivated by the fact that we can expect many bursts during the active periods of a source \citep{GerryZhang}. We simulate a 50 hour observation session spilt into 25 individual sessions of 2 hours each. Each 2 hour observing session is separated from the next one by a randomly chosen time gap lasting anywhere between a week and a year to resemble a real world scenario. The average wait time before the first repeat pulse is estimated for various combinations of $\gamma$ and rates as shown in Figure \ref{fig:weibull}.  

For a rate of 5.7 pulses day$^{-1}$ (middle panels of Figure \ref{fig:weibull}), we expect to wait an average of 30 minutes before the detection a repeat pulse in a 2 hour observing slot at Arecibo, at the redshift of the repeater. This is consistent with published observations of repeat pulses from FRB 121102 \citep{nat_spitler,Chatterjee,GerryZhang}. We expect similar wait-times for MeerKAT given the specifications in Table \ref{tab:specs}. From Figure \ref{fig:weibull}, on average we do not expect to be able to detect repeat pulses from FRBs originating at $ z \geq 0.5$ at both Arecibo and MeerKAT given the simulated observation duration and energy distribution. Similarly, the Parkes radio telescope is only capable of detecting repeat pulses from low redshift $z \leq 0.1$ FRB sources given the modelled observation time and repeat rate. The non-detection of a repeat pulse from FRB 010724 despite having spent $> 200$ hours on follow-up observations \citep{Rane} rules out the possibility of it repeating with the same rate as FRB 121102.
% This is consistent with not having detected a repeat pulse from FRB 010724 despite having spent $> 200$ hours on follow-up observations, with the major caveat that its repeat rate is the same as FRB 121102. 

Of the 23 published ASKAP FRBs, the one with the lowest DM (FRB 171020) corresponds to a maximum redshift of $z = 0.08$ \citep{nat_Shannon, Mahony}. Our simulations of the wait-times show that it takes $\sim 48$ hours on average to detect a repeat pulse at $z = 0.1$. \cite{nat_Shannon} have spent 185 - 1,097 hours following up the FRB positions after their initial detections to search for repeats and detected none. The non-detection of a repeat pulse from the ASKAP FRB 171020 based on our estimated average repeat time indicates that it does not possess the same rate statistics as FRB 121102, which is consistent with the conclusions of \cite{nat_Shannon}.

The upper and lower panels of Figure \ref{fig:weibull} represent rates of 1 pulse day$^{-1}$ and 10 pulses day$^{-1}$. We are also able to rule out a rate of 10 pulses day$^{-1}$ for FRBs 010724 and 171020 based on the non-detection of repeat pulses in the large amount of time spent on follow-up observations.

\section{Discussion}
\label{sec:disc}

It is evident from the simulations of both the Poisson and Weibull distribution models, that if the slope of the energy distribution is steep  (e.g. $\gamma = -1.7$), detectability of repeat pulses can be increased by performing follow-up observations 
of bright, low redshift FRBs. A major caveat to this is that it is highly dependent on the sensitivity of the instrument being used for the observations (see Figure \ref{fig:obstime}). 
% For instance, simulated FRBs at the Parkes radio telescope would only be detectable as repeaters provided $\beta \leq 3$ seconds. 
\cite{Hardy} report 13 radio pulses from FRB 121102 detected at the Effelsberg telescope of which two pulses were separated by only 34 ms. For an energy distribution slope of $\gamma = -1.7$, an observed wait time of 34 ms would imply a much shorter intrinsic wait timescale. While the observed timescale might help constrain periodicity, it does not rule out the possibility of emission in multiple rotational phase windows of a longer period \citep{Hardy}. Estimating an underlying wait time or periodicity with only a handful of bursts is exceedingly difficult. 
Based on this result, we suggest following up FRB positions with telescopes which are much more sensitive than the detection telescope. 
% The distribution of one-off events as a function of redshift is shown in Figure \ref{fig:waittime}. We see that more sensitive instruments like Arecibo and MeerKAT could well detect pulses up to the maximum redshift simulated under the assumptions modelled. 

Our simulations cannot rule out the possibility of a single population of FRBs given the wait time distribution modelled.
If the statistics of the repeat rate were non-Poissonian, then initial bursts from FRBs are expected to be promptly accompanied by
additional bursts. In this case immediate follow-up observations are more likely to yield repeat bursts. Since the detectability of pulses following a Poisson process is only affected by the total time spent on source, we strongly recommend multiple short follow-up observations rather than long ones, in order to maximize the detectability of repeat pulses irrespective of whether the energy distribution is Poissonian or not. 

The simulations effectively assume standard candles. In the all-sky survey of 10,000 repeating FRBs of energies in the range $10^{28} - 10^{36}$ J with slope $\gamma = -1.7$, only 15 FRBs are detected as singular events. For a steep slope of the energy distribution as assumed here, the number of detectable pulses per repeating FRB source decreases with distance despite the volumetric increase in number of FRB sources, thereby resulting in detections of one-off events. Along with the combination of the randomly chosen position in the modelled beam pattern, the one-off events potentially correspond to the bright tail-end of the adopted energy distribution.

The integral source count distribution or the detected brightness distribution of FRBs is defined as,
$N_\mathrm{FRB}(>\mathcal{F}_\mathrm{min}) \propto \mathcal{F}_\mathrm{min}^{\alpha}$
where $\mathcal{F}_\mathrm{min}$ is the minimum detectable fluence in Jy ms and is telescope specific.
Figure \ref{fig:rates} shows the rates at various present and upcoming transient search programs at different telescopes such as Apertif/Westerbork Synthesis Radio Telescope \citep[$S_\mathrm{min} = 0.46$ Jy, FoV = $8.7$ deg$^{2}$;][]{YogeshM}, ASKAP \citep[$S_\mathrm{min} = 6.6$ Jy, FoV = $30$ deg$^{2}$;][]{Bannister} and MeerKAT \citep[$S_\mathrm{min} = 0.44$ Jy, FoV = $1.27$ deg$^{2}$;][]{meertrap, meertime} as a function of the slope of the log$N$-log$\mathcal{F}$ curve. The sensitivities of the telescopes are calculated for a 10$\sigma$ event of 1-ms duration. All telescopes have been normalised to the discovery rate of 0.0625 events d$^{-1}$ \citep{Bhandari} at 1.4 GHz achieved at the Parkes radio telescope. In general, flatter slopes are seen to favour less sensitive, larger field-of-view telescopes while steeper slopes are seen to favour more sensitive, smaller field-of-view telescopes \citep{Trott, Hassall}.

The MeerTRAP project at the MeerKAT radio interferometer in South Africa will undertake high time resolution, fully commensal transient searches in parallel with all the MeerKAT Large Survey 
Project observations (MLSPs) resulting in hundreds of hours of on-sky time over the next few years \citep{meertrap}.
MeerTRAP will continuously and simultaneously use both the coherent and incoherent modes to probe two different parts of the FRB luminosity function. The incoherent mode will be more sensitive to the closer, brighter FRBs while the coherent mode will favour the distant and much fainter FRBs. From Figure \ref{fig:rates} we see that MeerTRAP is expected to detect at least $\sim 10$ FRBs a year irrespective of the slope of the integral source count distribution due to the simultaneous operation of the coherent and incoherent modes.
The precise localisation (few arc-seconds to sub-arcseconds) possible with MeerKAT either
through a detection in the narrow tied-array beams or in rapid imaging of buffered raw antenna data, will allow for a more targeted search in radio for repeats and in other 
wavelengths (e.g. optical with MeerLICHT) for afterglows \citep{meertrap}. As seen in Figure \ref{fig:obstime}, the sensitivity of the (coherent) MeerKAT array is only different by a factor of two from that of the Arecibo telescope ($\sim 0.09$ Jy ms and $\sim 0.04$ Jy ms respectively for a 10$\sigma$, 1 ms wide event), which makes it a very useful instrument for detecting repeating FRBs in addition to one-off events. 
It is possible that the presence of some external burst magnification mechanism (e.g. lensing) favours the repeated detection of FRB 121102 \citep{Cordes_lensing}. 
In any case, localisation along with the association of an FRB with a counterpart at another wavelength to determine the nature of the progenitor \citep[e.g. super-luminous supernovae;][]{Metzger} is key to resolving the existence of multiple populations. Detecting a typical wait timescale would also provide a constraint on possible populations.    

% \begin{itemize}
% \item Follow-up strategy: close bright ones (the probability of seeing a fainter one is higher than a brighter one for a steep slope. But this again depends on if your instrument is sensitive enough to detect
% faint pulses)
% \item Localization is key. Helpful to determine the origin. Repeaters could be SNe and non-repeaters could be AGN. Both could be in similar host galaxies so the actual source is important.
% \item It is possible that the presence of some external burst magnification mechanism (e.g. lensing) is permitting the repeated detection of FRB 121102
% \item statistics and repeat rates of FRB pulses could vary from source to source even if they are from a single progenitor class. 121102 may not provide global constraints
% \item In case of a Poissonian repetition, detection statistics are affected only by total time on source. For pulses clustered in time, shorter observations spanning the length of a single long observation are
% more likely to yield detections
% \item If the statistics of the repeat rate were non-Poissonian then initial bursts from FRBs are expected to have aftershocks similar to earthquakes. In this case immediate follow-up observations are more likely to yield repeats bursts
% \item Need multi-wavelength observations.
% \item Comment on multiple-populations??
% \end{itemize}

\section{conclusions}
\label{sec:conc}

In this paper we discuss the possibility of all FRBs being repeating sources similar to FRB 121102 with the telescope sensitivity and the time spent following up a field to look for repeats being the two major reasons for the observed dichotomy. The spatial number density of the simulated repeating FRBs is chosen to closely follow the cosmic SFH out to $z \sim 5$ since most FRB progenitor models (with the exception of the double compact merger model where there is a delay between the star formation and merger) are expected to track the star formation rate. Each repeater generates pulses 
with energies sampled from a power law energy distribution of slope $\gamma$, and the DM contribution to each repeater arises from the ISM of a putative host galaxy, the ISM of Galaxy and the IGM.
The time intervals between repeat pulses are chosen from a Poissonian random exponential with a rate parameter, $1 \leq \beta \leq 50$ seconds in one scenario, and from a Weibull distribution of arrival times with repetition rates of 1, 5.7 \citep{Oppermann} and 10 pulses day$^{-1}$ for a clustering parameter $k = 0.34$ \citep{Oppermann} in the other scenario.

Our simulations cannot rule out the possibility of a single FRB population given the energy distribution modelled. Comparisons of our simulated wait-times following a Weibull pulse arrival time distribution with real follow-up observations rule out the possibility of FRBs 171020 and 010724 repeating with the same rate as FRB 121102. We are also able to rule out a rate of 10 pulses day$^{-1}$.
% However, the differing pulse widths and spectra that have been observed between FRB pulses may provide modest support towards the existence of more than one population. The spectra of the repeater bursts are not well described by a typical power law and vary significantly
% from burst to burst indicating that the variations are likely intrinsic.
Similarly, based on comparisons between our simulations and follow-up observations of FRBs 010724 and 150708 at the Parkes radio telescope, we constrain the slope of the intrinsic energy distribution to be $-2.0 < \gamma < -1.0$. Irrespective of whether the intrinsic energy distribution is a power-law or a Weibull, several short observations are more likely to detect a repeat pulse from an FRB source than a single observation of the same length. All-sky simulations of a population of repeating FRBs at Parkes suggest that the detection of one-off events correspond to the bright tail-end of the adopted energy distribution due to the combination of the increase in volume probed with distance, and the position of the burst in the telescope beam.
Future wide-band receivers with high sensitivities like Arecibo and MeerKAT would prove beneficial in detecting repeat pulses from FRB sources discovered by less sensitive telescopes like Parkes and ASKAP (incoherent). However detection of repeat pulses with less sensitive instruments like Parkes would provide strong constraints on the intrinsic repeat timescales and progenitors.

% \begin{itemize}
% \item We cannot rule out the possibility of a single population given the luminosity distribution modelled
% \item However, the differing pulse widths and spectra that have been observed may provide modest support. The spectra of the Arecibo bursts are not well described by a typical power law and vary significantly
% from burst to burst. The variations are likely intrinsic.
% \item In any case several short observations are more likely to detect an FRB than a single
% observation of the same length as evident from observations of FRB 121102.
% \item luminosity function is constrained to $-2 < \gamma < -1.3$
% \end{itemize}

\section*{Acknowledgements}
The authors would like to thank the anonymous referee for their valuable comments. MC would like to thank Jason Hessels, Benito Marcote and Daniele Michilli for sharing the Arecibo PUPPI data for FRB 121102. MC would like to thank Evan Keane, Vincent Morello and Vivek Venkatraman Krishnan for useful discussions. The authors acknowledge funding from the European Research Council (ERC) under the European Union's Horizon 2020 research and innovation programme (grant agreement No 694745). Parts of this
research were conducted by the Australian Research Council Centre
for All-Sky Astrophysics (CAASTRO), through project number
CE110001020.
This work was performed on the gSTAR national facility at Swinburne University of Technology. gSTAR is funded by Swinburne and the Australian Government's Education Investment Fund.

%----------------------------------------------------------------------------------------
% BIBLIOGRAPHY 
%----------------------------------------------------------------------------------------
\bibliographystyle{mnras}
\bibliography{sims}

% Don't change these lines
\bsp	% typesetting comment
\label{lastpage}
\end{document}